\documentclass[conference]{IEEEtran}
\IEEEoverridecommandlockouts
\usepackage{amsmath,amssymb,amsfonts}
\usepackage{algorithmic}
\usepackage{enumitem}
\usepackage{subfigure}
\usepackage{graphicx}
\usepackage{textcomp}
\usepackage{gensymb}
\usepackage{xcolor}
\usepackage{soul}
\usepackage{blindtext, graphicx}
\usepackage{tabularx}
\usepackage{multicol}
 \usepackage{relsize}
 \usepackage{stfloats}
\usepackage{cuted}
\usepackage{cite}
\usepackage{url}
\usepackage{multirow}
\usepackage{hyperref}
\usepackage[cmintegrals]{newtxmath}
\usepackage{graphics}
\usepackage{float}

\usepackage{cite}
\usepackage{booktabs}
\usepackage[normalem]{ulem}

\def\BibTeX{{\rm B\kern-.05em{\sc i\kern-.025em b}\kern-.08em
    T\kern-.1667em\lower.7ex\hbox{E}\kern-.125emX}}
\begin{document}

\title{Analysis and Empirical Validation of \\ Visible Light Path Loss Model for Vehicular Sensing and Communication
\thanks{This work was supported by U.S. Department of Transportation through the Transportation Consortium of South-Central States (Tran-SET) under Grant No. 18ITSOKS01.}

\thanks{This work has been submitted to the IEEE for possible publication.
Copyright may be transferred without notice, after which this version may
no longer be accessible.}
}

\author{
\IEEEauthorblockN{Hisham Abuella}
\IEEEauthorblockA{\textit{School of Electrical and Computer}\\
\textit{Engineering}\\
\textit{Oklahoma State University}\\
Stillwater, OK, USA \\
hisham.abuella@okstate.edu}
\and
\IEEEauthorblockN{Md Zobaer Islam}
\IEEEauthorblockA{\textit{School of Electrical and Computer}\\
\textit{Engineering}\\
\textit{Oklahoma State University}\\
Stillwater, OK, USA \\
zobaer.islam@@okstate.edu}
\and
\IEEEauthorblockN{Russ Messenger}
\IEEEauthorblockA{\textit{School of Electrical and Computer}\\
\textit{Engineering}\\
\textit{Oklahoma State University}\\
Stillwater, OK, USA \\
russ.messenger@okstate.edu}
\and
\hspace{3cm}
\IEEEauthorblockN{John F. O'Hara}
\IEEEauthorblockA{\hspace{3cm}\textit{School of Electrical and Computer}\\
\textit{\hspace{3cm}Engineering}\\
\textit{\hspace{3cm}Oklahoma State University}\\
\hspace{3cm}Stillwater, OK, USA \\
\hspace{3cm}oharaj@okstate.edu}
\and
\hspace{-1.1cm}
\IEEEauthorblockN{Sabit Ekin}
\IEEEauthorblockA{\hspace{-1cm}\textit{Department of Engineering Technology}\\
\textit{\hspace{-1cm}and Industrial Distribution}\\
\textit{\hspace{-1cm}Texas A\&M University}\\
\hspace{-1cm}College Station, TX, USA \\
\hspace{-1cm}sabitekin@tamu.edu}
}



\maketitle
\begin{abstract}

Advancements in lighting systems and photodetectors provide opportunities to develop viable alternatives to conventional communication and sensing technologies, especially in the vehicular industry. Most of the studies that propose visible light in communication or sensing adopt the Lambertian propagation (path loss) model. This model requires knowledge and utilization of multiple parameters to calculate the path loss such as photodetector area, incidence angle, and distance between transmitter and receiver. In this letter, a simplified path loss model that is mathematically more tractable is proposed for vehicular sensing and communication systems that use visible light technology. Field measurement campaigns are conducted to validate the performance and limits of the developed path loss model. The proposed model is used to fit the data collected at different ranges of incident angles and distances. Further, this model can be used for designing visible light-based communication and sensing systems to minimize the complexity of the Lambertian path loss model, particularly for cases where the incident angle between transmitter and receiver is relatively small.

\end{abstract}


%
\IEEEpeerreviewmaketitle

\section{Introduction}

Wireless communication and sensing systems have experienced many advancements, from the discovery of electromagnetic (EM) waves, to wireless telegraphs and radios, to modern smartphones, connected vehicles and Internet of Things (IoT) devices. All of these new technologies need wireless communication to adapt to the demand for higher bandwidth and to the inflation of data consumption. New technologies increasingly adopt wireless communications with ever higher bandwidth and data consumption to accomplish their purpose.  In some systems, like in the vehicular industry, utilizing the radio frequency (RF) technology can be inefficient due to interference, spectrum scarcity, health concerns and power limitations~\cite{VLC_State_of_Art,pathak2015visible}. In recent years, the interest for use of visible light systems to enable wireless connectivity in vehicle-to-vehicle (V2V) and vehicle-to-infrastructure (V2I) links has increased. It is also called as V2X communication. This increasing interest is due to the advancements in light-emitting diode (LED) and photodetector (PD) technologies. Moreover, these systems are  less costly, highly efficient and can support high data rates. One important requirement for visible light communication is to avoid human perception of flickering at the light source, an easily achievable task by employing intensity modulation faster than 200 Hz~\cite{IEEE_802_15_7_VLC}. 

There are multiple studies discussing different implementations of visible light communication (VLC) in V2X applications. Lui \textit{et al.} introduced the idea of enabling Vehicular VLC (V2LC) systems and studied the effect of multiple vehicles and the visible light noise and interference~\cite{V2V_VLC_Ref1}. They considered the reliability and latency requirements to examine the V2LC performance. Multiple studies have demonstrated the VLC systems for outdoor vehicular scenarios. Cailean \textit{et al.} presented a short distance prototype to transmit data using Miller and Manchester coding~\cite{V2V_VLC_Ref2}. Lourenco \textit{et al.} discussed the challenges that outdoor VLC systems face and implemented a prototype that can achieve low data rates in presence of high optical noise levels~\cite{V2V_VLC_Ref3}. 

In most of the previously mentioned studies, understating the path loss model and estimating the received light power in different positions (different distances) are critical in estimating the theoretical performance limits of VLC systems. 
Few researchers have studied and analyzed the visible light channel and its path loss model behavior in real world scenarios, which are important and critical parameters to be considered. The Lambertian propagation (path loss) model was adopted by most of the VLC and VLS (visible light sensing) studies~\cite{VLC_State_of_Art,pathak2015visible,ChannelModel_survey,Vildar_Journal}. 
This propagation model works well in indoor scenarios. However, the same model may not work well for visible light communication and sensing in V2X applications. 

As a potential solution, some studies introduced empirical path loss models for the outdoor vehicular VLC systems. Cui \textit{et al.} studied an outdoor VLC link using LED traffic lights where the link path loss model is analyzed both theoretically and experimentally~\cite{Channel_VLC_Ref1}. Viriyasitavat \textit{et al.} derived a realistic path loss model for VLC systems using off-the-shelf scooter taillights. The proposed model accurately estimated the received power from the taillight up to 10 meters~\cite{Channel_VLC_Ref2}. Turan \textit{et al.} did a frequency domain channel sounding and characterization for a vehicular VLC in different scenarios~\cite{Channel_VLC_Ref4}. A comparison between radio frequency and visible light propagation channels for vehicular communication was presented by Cheng \textit{et al.} in~\cite{Channel_VLC_Ref5}. Recently, there is a shift in the studies to simplify the visible light path loss model for outdoor vehicular scenarios and to verify these models using simulations in ray tracing software, as done in~\cite{Ray_tracing_Eldeeb_Uysal_Conf_2019}. Moreover, Elamassie \textit{et al.} developed a path loss model for V2V links as a function of distance under different weather conditions and confirmed their models using ray tracing software~\cite{Elamassie_Channel_model_conf}. In addition to ray tracing tools, Eso \textit{et al.} performed an experimental investigation on the effects of fog on optical camera based VLC. Memedi \textit{et al.} investigated the impact of vehicle type and headlight characteristics on the VLC performance~\cite{Channel_VLC_Ref3}.
However, in the aforementioned studies, the proposed models are experimentally tested with only a limited number of static points (2D or 3D) between the receiver and transmitter.  

In this letter, a simplified visible light path loss model is proposed, analyzed and empirically validated for outdoor V2I communication and sensing applications. To verify the proposed model, data is collected from field measurements using off-the-shelf LED lights and a PD using a new dynamic channel modeling approach. This model can be used by different studies to decrease the analysis complexity (i.e., increase mathematical tractability), where the Lambertian propagation model would be assumed and the incident angle is small enough. Moreover, the limits of the developed path loss model are provided.

In summary, the main contributions of this letter are as follows:
\begin{itemize}
    \item Proposing and analyzing a mathematically more tractable and simplified visible light path loss model.
    \item Verifying the proposed path loss model by field measurements using a dynamic channel modeling approach.
    \item Discussing the limits of the proposed path loss model and dynamic channel modeling approach.
\end{itemize} 

As per our knowledge, this is the first work to provide and verify a simplified visible light path loss model for V2X applications that takes the motion of the vehicle (dynamic channel modeling) into account.

The rest of the letter is organized as follows. In Section~\ref{Sec:System_model}, the system model and the experimental setting are provided. In Section~\ref{Sec:Channel_Model},  the developed path loss model and its mathematical proof are given. In Section~\ref{Sec:Results}, the comparison and the field measurements are presented. Finally, the conclusions are discussed in Section~\ref{Sec:Conclusion}. 

\section{System model and experimental setting}
\label{Sec:System_model}
\begin{figure}[t]
\centering
\includegraphics[width=0.48\textwidth]{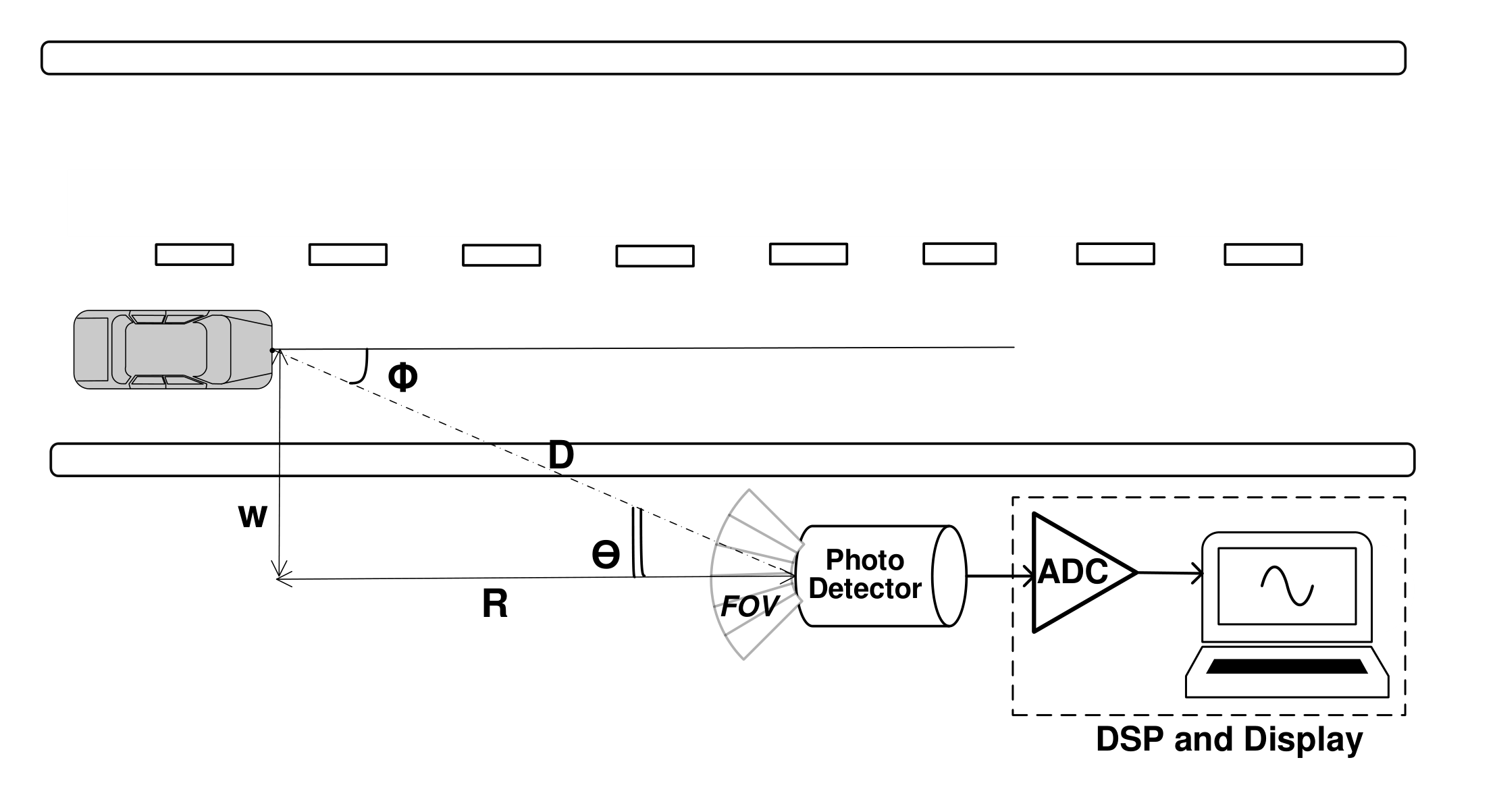}
\caption{The system model that is used in the study.}
\label{fig:System_model}
\end{figure}

The system model under consideration is shown in the Fig.~\ref{fig:System_model}, where $\theta$ and $w$ are the incidence angle and the lateral distance between the vehicle and the photodetector (PD), respectively. $R$ and $D$ are the varying longitudinal distance and the actual distance between the vehicle and the PD, respectively. For simplicity, longitudinal distance ($R$) will be referred to as range and actual distance between the vehicle and the PD ($D$) will be referred to as distance between the PD and vehicle for the rest of the paper.

\begin{figure}[t]
\includegraphics[width=0.48\textwidth]{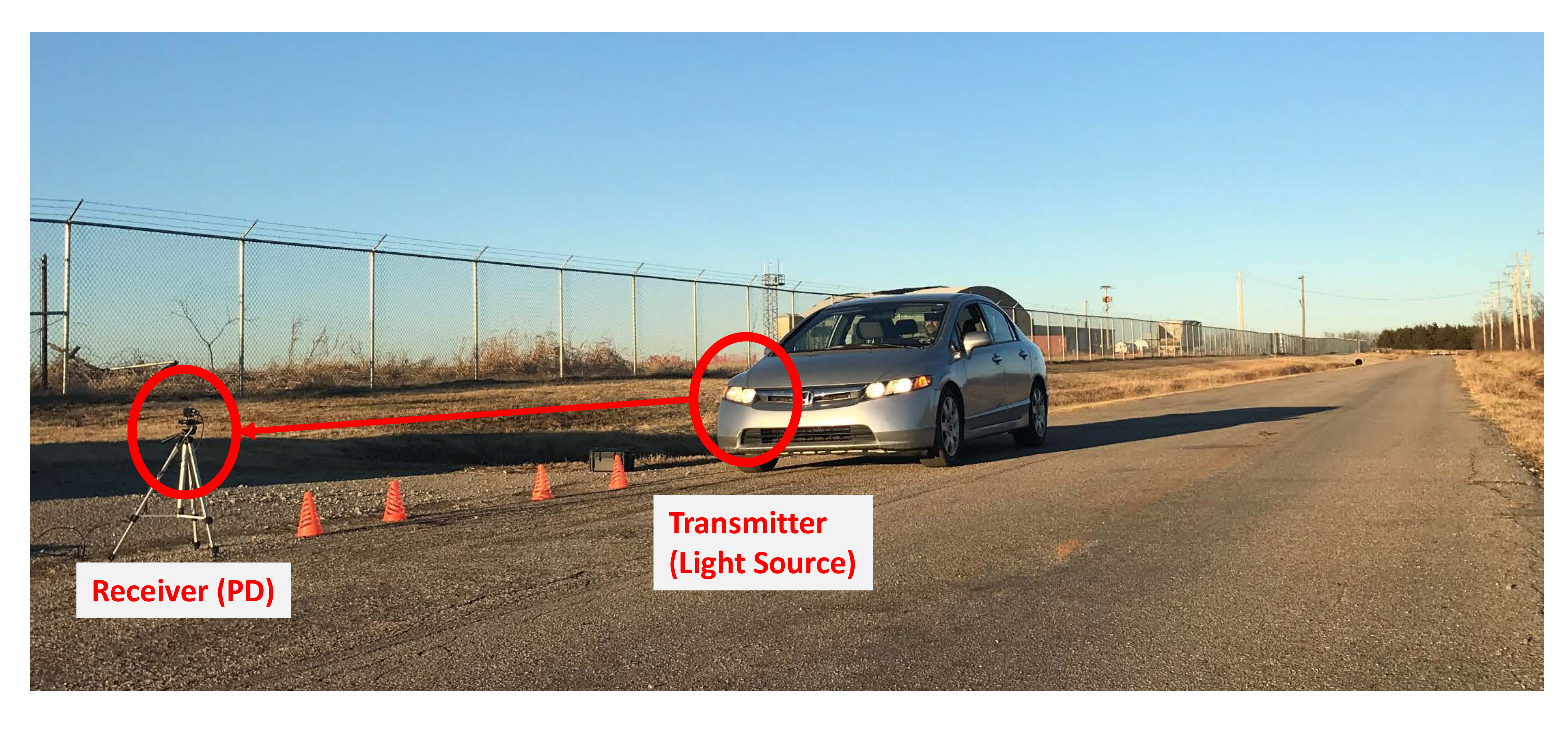} 
\centering
\caption{The experimental setting used for collecting data.}
\label{fig:experimental_setting}
\end{figure} 


The actual experimental setting used to collect the data is presented in Fig.~\ref{fig:experimental_setting}. A Honda Civic 2008 with a LASFIT L1 LED headlamp~\cite{LASFIT_LED} is used as the transmitter. A photodetector from Thorlabs (PDA 100A)~\cite{PDA_100A} is used as the receiver and a RaspberryPi Model3B+~\cite{RaspeberryPi} is used as the processing module. A PiPlate (DAQC2plate)~\cite{PiPlate} is utilized as an analog-to-digital converter (ADC) in the proposed V2X system. All the parameters used in the experimental setting are stated in Table~\ref{Experimental_Parameters}. A 50$\Omega$ BNC terminator is added to the PDA 100A which, according to the datasheet of the photodetector in~\cite{PDA_100A}, decreases the reflections and improves the signal to noise ratio of the collected signal which will be between 0V-5V. In addition, the received light power will be a function of this collected signal and can be converted according to the amplifier mode and the wavelength of the visible light used~\cite{PDA_100A}. A 335nm-610\,nm bandpass optical filter from Thorlabs is used to remove the effect of IR and lower wavelengths. The equation used to convert the received signal to the received light power is as follows~\cite{PDA_100A}:
  \begin{equation}
     \label{eq:PD_Power_Converstion}
                   P_{in} = \frac{2 V_{out}}{R_{PDA100A}(\lambda) G(G_{amp})},
  \end{equation}
where $P_{in}$ is the received light power at the PD (watts), $V_{out}$ is the output signal from the ADC (V), $R_{PDA100A}(\lambda)$ is the PDA100A responsivity to different wavelengths which is 0.4\,Amp/watt for 610\,nm wavelength, and $G(G_{amp})$ is the transimpedance gain (V/Amp) which will vary according to the chosen amplifier variable gain ($G_{amp}$) (from 0\,dB to 70\,dB).
Equation \eqref{eq:PD_Power_Converstion} according to~\cite{PDA_100A} can be simplified as:
\begin{equation}
     \label{eq:Final_PD_Power_Converstion}
                   [P_{in}]_{dBW} = 10\log_{10}(V_{out}) - 0.5G_{amp} - 21.76 ,
  \end{equation}
 where $[P_{in}]_{dBW}$ is the received light power at the PD in dBW scale.

\begin{table}[]
\renewcommand{\arraystretch}{1.15}
\caption{Parameters used in data collection experimental setting}
\label{Experimental_Parameters}
\begin{tabular}{@{}ll@{}}
\toprule
\textbf{Parameter}                  & \textbf{Value }                                \\  \midrule
 \midrule
Transmitter (Tx) LED            & Honda Civic 2008 with a LASFIT \\
                                 &       (L1 9005) LED         \\
Receiver (Rx) PD                & Thorlabs PDA 100A                     \\
Receiver detection area    & 10\,mm$^2$               \\
Avg. Tx power (electrical) & 20\,Watts                                   \\
Tx and Rx Height                     &  60\,cm                                     \\
ADC (PiPlate (DAQC2plate)      &    366\,$\mu V$ per bit, up to 12\,$V$ input range, and \\ 
      &    50\,KHz Samples/sec   \\
\bottomrule
\end{tabular}
\end{table}

A Python script is used to collect the data samples and save them on the RaspberryPi. Finally, offline data processing is done on a laptop using MATLAB scripts.

\section{Proposed Channel Model}
\label{Sec:Channel_Model}

 The visible light channel has been studied intensively in the context of indoor communications~\cite{VLC_State_of_Art,pathak2015visible}. One of the most used and adopted line-of-sight (LOS) channel models is Lambertian model~\cite{VLC_First_channel_Model}. 
 The Lambertian model presents the effects of different variables and parameters that make the visible light signal vary at the receiver end. The parameters are the transmitter power, the distance between light source and PD, optical PD size, PD field of view (FOV) (which is the region of space where the detector can detect any light entering it) and the incident angle ($\theta$). The Lambertian model for the received signal power-distance relation is given as follows:
   \begin{equation}
   \label{eq:Lambertian_Channel1}
                  P_r=\frac{(n+1)A_R P_t}{2\pi D^{\gamma}}\cos^n(\phi)\cos(\theta) , \forall\theta< \phi_{1/2},
  \end{equation}
where $P_t$ is the transmitter power in watts, $P_r$ is the received signal power in watts and $A_R$ is the optical detector size.  $ \phi $ and $ \theta $ are angles of irradiance and incidence, respectively. The path loss exponent $\gamma$ depends on the environmental conditions, such as reflectiveness of materials, light conditions, etc. Typically, the range of path loss exponent lies between 1 and 6. 
 
In addition, $\phi_{1/2}$ is the semi-angle at half-power of the LED (which is half of the FOV of the light source), and $n$ is the order of the Lambertian model and is given by
\begin{equation}
n=-\frac{\ln(2)}{\ln(\cos\phi_{1/2})}.
\end{equation}

In our case, the light source and the receiver have the same movement directions and at the same height, therefore,
     \begin{equation}
     \label{eq:Theta_PHI}
                   \theta = \phi,
  \end{equation}
 where  $ 0 <\theta < \phi_{1/2} $. Using~\eqref{eq:Theta_PHI},~\eqref{eq:Lambertian_Channel1} can be further simplified as
\begin{equation}
 \label{eq:Lambertian_Channel2}
 P_r=\frac{(n+1)A_R P_t}{2\pi D^{\gamma}}\cos^{n+1}(\theta).
 \end{equation}
 
Then, in order to derive $P_r(t)$ in terms of $D(t)$,~\eqref{eq:Lambertian_Channel2} is further simplified by defining a constant $K$ as  
\begin{equation}
K = \frac{(n+1)A_R P_t}{2\pi}.
\end{equation}
Which leads to, 
\begin{equation}
 \label{eq:Lambertian_Channel3}
 P_r=K D^{- \gamma}\cos^{n+1}(\theta).
 \end{equation}
 Taking $10 \log_{10}(.)$ at both sides,
 
 \begin{equation}
 \label{eq:Lambertian_Channel_Log}
 P_{r_{dB}}=K_{dB} - \gamma D_{dB} + 10(n+1) \log_{10}(\cos(\theta)),
 \end{equation}
where $P_{r_{dB}} = 10 \log_{10}(P_{r})$, $K_{dB} = 10 \log_{10}(K)$, and $D_{dB} = 10 \log_{10}(D)$.

It is given that $\cos(\theta)=\frac{R}{D}=\frac{\sqrt{D^2-w^2}}{D}=\sqrt{1-\frac{w^2}{D^2}}$.
Hence, 
 \begin{equation}
 \label{eq:Lambertian_Channel_Log_final}
 P_{r_{dB}}=K_{dB} - \gamma D_{dB} + 5(n+1) \log_{10}\left( 1-\frac{w^2}{D^2}\right ).
 \end{equation}

Finally, the expression can be divided into two conditions:
 \begin{equation}
 \label{eq:conditiona_eq}
P_{r_{dB}}=\begin{cases}K_{dB} - \gamma D_{dB},&\frac{w^2}{D^2} << 1
\\K_{dB} - \gamma D_{dB} + G_{dB},&otherwise \end{cases}
 \end{equation}
where $G_{dB}= 5(n+1) \log_{10}\left(1-\frac{w^2}{D^2}\right)$. The first condition (far-scenario) is when $D$ is large enough or when $w$ is small, which can also be called as log-linear simplified model as $P_{r_{dB}}=K_{dB} - \gamma D_{dB}$ $\rightarrow$  $P_r=K D^{-\gamma}$.  While the second condition (near-scenario) is when $D$ is small or when $w$ is large compared to $D$.

Notice that for a constant value of $K_{dB}$ and $\gamma$, the received power changes with the distance. The longer the distance between the transmitter (light source, e.g. LED) and receiver (PD), the lower the received power is and a factor $G_{dB}$ is added in near-scenarios.

\section{Measurement Results and Verification}
\label{Sec:Results}
In this section, the data collected from the setting presented in Fig.~\ref{fig:experimental_setting} in both static and dynamic channel modeling scenarios are presented. In case of dynamic channel modeling, data for both night and daylight environments is presented.

\subsection{Static Channel Modeling Scenario}
\begin{figure}[t]
\centering
\includegraphics[width=0.49\textwidth]{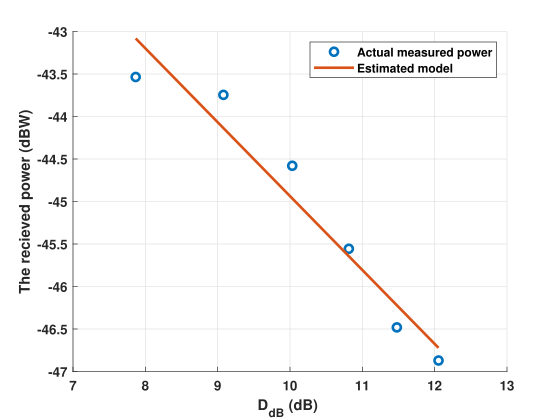}
\caption{Static channel model with vehicle equipped with LED headlamps (night scenario measurements).}
\label{fig:Static_Channel_Model}
\end{figure}

In Fig.~\ref{fig:Static_Channel_Model}, the channel model is estimated after measuring the received power at several distance points at night and fitting the collected measurement values to estimate the best channel model parameters to fit the data. As shown in Fig.~\ref{fig:Static_Channel_Model}, the data is collected on multiple distances between (8\,m-12\,m) with 200 samples averaged at each point (2 seconds of data at 100 samples/sec sampling rate). As the distance increased, the received power decreased at an exponential rate of $\gamma$. It is evident that the measurements at different distance points do not provide a complete picture of how the channel model is changing and behaving over different regions explained in Section~\ref{Sec:Channel_Model}. However one can observe that  the linearity of the figure starts decreasing when $D_{dB}< 10$. Therefore, continuous measurements while vehicle approaching towards the PD would provide a better picture for realistic visible light channel modeling, which is discussed in the next section.


\subsection{Dynamic Channel Modeling Scenario}
The dynamic channel modeling is followed to overcome the limitations discussed in the previous subsection. We will first present how dynamic channel modeling is performed and the assumptions are taken when performing it. Then, the data collected at night and in sunny daylight conditions are presented.

\begin{figure}[t]
\centering
\includegraphics[width=0.48\textwidth]{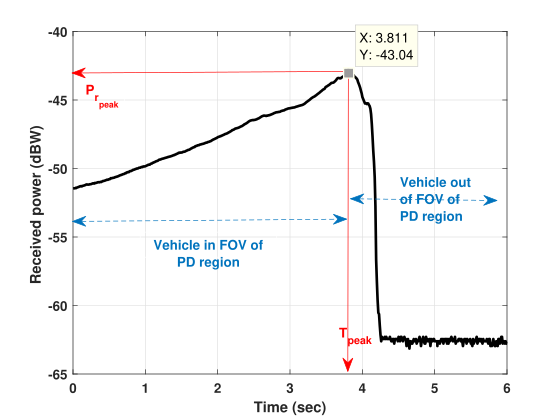}
\caption{Received power (dB) at night versus time (sec) with vehicle equipped with normal headlamps.}
\label{fig:Time_Power_Peak}
\end{figure}

The received light power at the PD is collected for a vehicle approaching the PD with a constant speed ($V$) of 20\,mph (8.9408\,m/sec) as shown in Fig.~\ref{fig:System_model}. The data is collected for 10 seconds with a sampling rate of 100 samples/sec (total of 1000 samples). After identifying the range (R) where the peak of the received power happens ($R_{peak}$) and saving it as a reference, the peak of the data saved is located as shown in Fig.~\ref{fig:Time_Power_Peak}. Then, the curve is flipped and the power value corresponding to each distance point is identified using the timing information. Therefore, the range (R) of each data point is calculated as follows:
 \begin{equation}
 \label{eq:Dynamic_Channel_Model}
  R_i = R_{Peak} + V (T_{Peak} - t_i),
 \end{equation}
where $R_i$ is the range (horizontal distance between vehicle and PD, $T_{Peak}$ time stamp of the data point where the peak is measured, and $t_i$ is the time stamp of each data point measured.

Additional details for the channel modeling experimental setup are as follows:

\begin{itemize}
\item The vehicle is moving at a known constant speed ($V$).

\item Vertical distance of the vehicle from PD ($w$) is also constant and known.

\item The range (R) at which the peak power is received at PD is known.

\end{itemize}

After transforming the time axis with range (R) or actual distance (D) axis using~\eqref{eq:Dynamic_Channel_Model}, the channel model can be estimated using linear fitting as shown in $P_{r_{dB}}= K_{dB} - \gamma D_{dB}$. 



\begin{table}[]
\renewcommand{\arraystretch}{1.4}
\centering
\caption{Channel parameters estimated using dynamic and static channel modeling in night and daylight environments.}
\label{tab:Dynamic_channel_parameters}
\begin{tabular}{|c|c|c|}
\hline
\textbf{Environment }    & \textbf{$K_{dB}$} & \textbf{$\gamma$} \\ \hline
Night  vehicle&   -35.2680\,dB      &   0.9707                  \\ \hline
Sunny daylight  vehicle&  -32.6335\,dB       & 0.0175                  \\ \hline
\end{tabular}
\end{table}

\begin{figure}[t]
\centering
\includegraphics[width=0.49\textwidth]{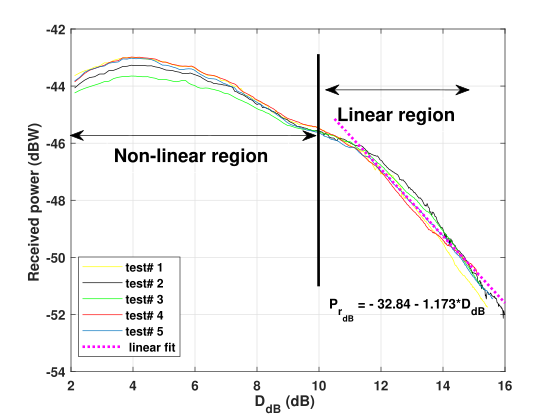}
\caption{Received power (dBW) at night versus distance in log domain (dB) for different test cases and expression for linear region with  $K_{dB}=-32.84$ and $\gamma=1.173$.}

\label{fig:Normal_Car_Night_Log_Plot_2_headlamps}
\end{figure}

\begin{figure}[t]
\centering
\includegraphics[width=0.49\textwidth]{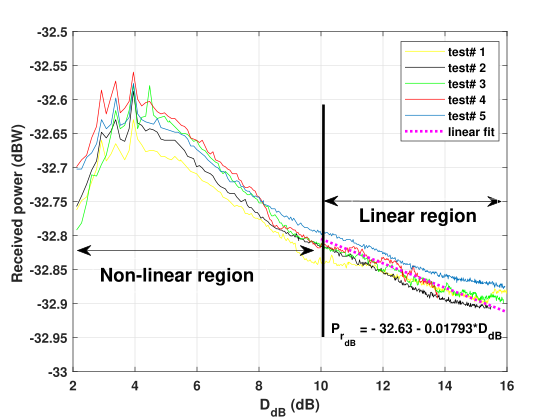}
\caption{Received power (dBW) at sunny daylight versus distance in log domain (dB) for different test cases and expression for linear region with  $K_{dB}=-32.63$ and $\gamma=-0.01793$.}

\label{fig:Normal__Car_Sunnyday_Log_Plot_2_headlamps}
\end{figure}

As shown in Fig.~\ref{fig:Normal_Car_Night_Log_Plot_2_headlamps}, it is clear that the linear region of the channel model starts at a range larger than around 10 meters. When the range is less, the channel behavior is seen to be non-linear which is because of $G_{dB}$ in~\eqref{eq:conditiona_eq} (when condition $\frac{w^2}{D^2} << 1$ is not satisfied). In addition, it is clear that there is a difference in received light power at night and sunny daylight scenarios as shown in Fig.~\ref{fig:Normal__Car_Sunnyday_Log_Plot_2_headlamps}. In daylight scenario, power received by the PD is higher and noisier because of the additional interference from sunlight.

\section{Conclusions}
\label{Sec:Conclusion}
This letter analyzed and validated a mathematically more tractable path loss channel model that can be used for vehicular sensing and communication applications. According to the developed model, when the incident angle of light at the receiver was small, the received light power became linear with respect to logarithmic distance between the transmitter and the receiver. The model was validated by a set of experimental measurements in both static and dynamic scenarios, and in both night and daytime settings. This proposed model can be utilized in multiple visible light enabled sensing and communication applications where reduction of the complexity of the channel model is needed.

\bibliographystyle{IEEEtran}
\bibliography{ref.bib}






\end{document}